\documentclass[%
 preprint,
 amsmath,amssymb,
 aps,
]{revtex4-2}

\usepackage{graphicx}% Include figure files
\usepackage{dcolumn}% Align table columns on decimal point
\usepackage{bm}% bold math
\begin{document}

\title{Effective binding potential from Casimir interactions: the case of the Bose gas}

\author{Marcin Pruszczyk}
\email{m.pruszczyk2@student.uw.edu.pl}
\author{Pawe{\l} Jakubczyk}
 \email{pawel.jakubczyk@fuw.edu.pl}
\affiliation{Institute of Theoretical Physics, Faculty of Physics, University of Warsaw, Pasteura 5, 02-093 Warsaw, Poland
 }%

\date{\today}% It is always \today, today,
             %  but any date may be explicitly specified

\begin{abstract}
We consider the thermal Casimir effect in ideal Bose gases, where the dispersion relation involves both terms quadratic and quartic in momentum. We demonstrate that if macroscopic objects are immersed in such a fluid in spatial dimensionality $d\in\{3,7, 11, \dots\}$ and at the critical temperature $T_c$, the Casimir force acting between them is characterized by a sign which depends on the separation $D$ between the bodies and changes from attractive at large distances to repulsive at smaller separations. In consequence, an effective potential which binds the two objects at a finite separation arises. We demonstrate that for odd integer dimensionality $d\in \{3, 5, 7, \dots\}$, the Casimir energy is a polynomial of degree $(d-1)$ in $D^{-2}$. We point out a very special role of dimensionality $d=3$, where we derive a strikingly simple form of the Casimir energy as a function of $D$ at Bose-Einstein condensation. We discuss crossover between monotonous and oscillatory decay of the Casimir interaction above the condensation temperature.    
\end{abstract}

%\keywords{Suggested keywords}%Use showkeys class option if keyword
                              %display desired
\maketitle

%\tableofcontents

\section{Introduction}
The paradigm of fluctuation-induced, Casimir-type forces, was over the last decades invoked in a multitude of physical situations \cite{Krech_book, Kardar_1999, Brankov_book, Klimchitskaya_2009, Maciolek_2018}. The Casimir effect was first recognized \cite{Casimir_1948} as a prediction of quantum electrodynamics in 1948, and since then was discussed in contexts as diverse as classical fluids, Bose-Einstein condensates, biological cells, and cosmology with a variety of methods encompassing experimental (e.g. Refs.~\onlinecite{Lamoreaux_1997, Soyka_2008}), analytical (see \onlinecite{Dantchev_2022} for a recent review), and numerical (e.g. Refs.~\onlinecite{Vasilyev_2009, Hasenbush_2012, Hasenbush_2015}) approaches. In a general setting, one may characterize the Casimir effect as an effective, long-ranged force, which arises when macroscopic objects are placed in a system characterized by strong fluctuations.  The necessarily occurring boundary conditions at the surfaces of the bodies give rise to deformations of the spectrum of the fluctuations. A broad range of situations may be realized depending on the nature of the fluctuations (being it the ones occuring in e.g. quantum vacuum, classical fluid at criticality, or Bose-Einstein condensate), as well as the character of the boundary conditions imposed on the fluctuations by the immersed objects. Nonetheless, the Casimir effect is characterized by a high degree of universality. 

In typical situations, the Casimir force $F$ is attractive and decays as a power of the distance $D$, $F\sim D^{-\gamma}$. The question concerning systems where the Casimir interaction would appear repulsive received quite a substantial interest over the years. Proposals of such setups \cite{Soyka_2008, Nellen_2009, Gross_2016, Jakubczyk_2016_2, Flachi_2017, Faruk_2018, Jiang_2019}  involved \emph{inter alia} nonuniform media, asymmetric boundary conditions, or anisotropic properties. 

In a sequence of recent studies \cite{Burgsmuller_2010, Lebek_2020, Lebek_2021, Napiorkowski_2021}, it was demonstrated that the sign of the Casimir force is sensitive to the asymptotic, low-momentum behavior of the dispersion relation $\epsilon_{\vec{k}}$ of the medium. In particular, in dimensionality $d=3$ it becomes generally repulsive if the standard behavior $\epsilon_{\vec{k}}\sim k^2$ is replaced by a quartic form $\epsilon_{\vec{k}}\sim k^4$ in at least some of the directions of the wave vector $\vec{k}$. These finding are pertinent to lattice setups, where the single-particle dispersion may be tuned so that its  leading (generically quadratic) asymptotic term becomes tuned towards zero, and the usually subleading, quartic contribution becomes the dominant one \cite{Jakubczyk_2018}. The abovementioned earlier studies considered situations where the dispersion relation involved (in each direction) only one term, typically either quartic or quadratic in momentum. In the present paper we propose an extension, where both quadratic and quartic contributions are accounted for, and, by manipulating the dispersion parameters, one may evolve the system between configurations corresponding to attractive and repulsive Casimir interactions. Situations of this type arise in fact in physical lattice systems, where the dispersion relation involves terms of arbitrary order in momentum. In such cases it is justified to replace it with the leading low-energy asymptotics only, provided one is interested exclusively in the dominant, large distance behavior of the Casimir interaction.   

As a prototype system, we consider the ideal Bose gas equipped with a tunable dispersion, which allows, in a number of physically relevant cases, for a fully analytical resolution of the problem. The outline of the paper is as follows: in Sec.~II we present the model and the key steps of the solution; in Sec.~III we give our major results. Sec.~IV contains a discussion and summary. Details of the derivations are contained in the Appendix.         
\section{The model and its solution}
We consider a system of spinless bosonic particles in a $d$-dimensional hypercuboid vessel of sidelength $L$ in $d-1$ directions and sidelength $D$ in the $d$-th direction. Periodic  boundary conditions are imposed on the one-particle wave functions of the particles so that the wave vectors labeling single-particle momentum eigenstates of the system are of the form
\begin{equation}
\mathbf{k} = 2\pi\left(\frac{n_1}{L}, \frac{n_2}{L}, \ldots, \frac{n_{d-1}}{L}, \frac{n_d}{D} \right) 
\label{spectrum}
\end{equation}
for each $n_i \in \mathbb{Z}$ with $i \in \{1, 2, \ldots, d\}$.   	
The Hamiltonian governing the system reads
\begin{equation}
\hat{H}_0 = \sum_{\mathbf{k}}\varepsilon_{\mathbf{k}}\hat{n}_{\mathbf{k}}\;.
\end{equation}
Here the symbols $\{\hat{n}_{\mathbf{k}}\}$ denote the particle  number operators and the summation is performed over the one-particle states labeled by $\mathbf{k}$. We assume that the dispersion relation of the particles takes the following form:
\begin{equation}
\varepsilon_\mathbf{k} = t \sum_{i = 1}^{d-1}k_i^2 + \tau k_d^2 + \tau' k_d^4\;, \;\; t>0\;, \;\; \tau, \tau' \geq 0\;, \;\; \tau + \tau'>0\;, 
\label{dispersion}
\end{equation}
where $k_i$ is the $i$-th component of the wavevector. Such an asymptotic small-$|\mathbf{k}|$ form of the dispersion may be obtained in an anisotropic optical lattice as discussed in Ref.~\cite{Jakubczyk_2018} and is potentially experimentally realizable in optical lattice cold atom setups with tunable dispersions \cite{Greif_2013, Imriska_2014}.  The non-zero value of the parameter $\tau'$ is the key feature of the system. We will be particularly interested in the limiting situation, where the parameter $\tau$ is continuously tuned towards zero and the role of the term $\tau' k_d^4$ becomes of increasing importance. 

Our analysis employs the grand canonical framework. The grand canonical potential is decomposed as: 
\begin{equation}
	\Omega(T, \mu, L, D) = L^{d-1} D \omega_b(T, \mu) + L^{d-1}\omega_s(T, \mu, D) + \ldots\;, 
 \label{decomposition}
\end{equation}
where $\omega_b$ is the bulk contribution to the thermodynamic potential, $\omega_s$ is the surface contribution, while ``$\ldots$'' denotes contributions, which scale as lower powers of the linear system size and may be disregarded in the limits considered in the present study. The Casimir force is obtained as: 
\begin{equation}
	F_C(T, \mu, D) = -\frac{\partial\omega_s(T, \mu, D)}{\partial D}\;.
\end{equation} 
Our present strategy amounts to evaluating 
\begin{equation}
\omega_s(T, \mu, D) = \lim_{L\to\infty}\frac{1}{L^{d-1}}\left[\Omega(T,\mu, L,D)-L^{d-1}D\omega_b(T,\mu)\right]   \;, 
\end{equation}
where $\omega_b\left(T, \mu\right)$ follows from:
\begin{equation}
\omega_b\left(T, \mu\right) = \lim_{L\to\infty} \frac{1}{L^{d}}\Omega(T,\mu,L,D=L)\;.    
\end{equation}
Note that the $d-1$-dimensional ``walls'' between which the Casimir force acts are oriented perpendicular to the special $d$-th spatial direction [Eq.~(\ref{dispersion})].  

The grand canonical potential of the non-interacting bosonic system is given by \cite{Ziff_1977}
\begin{equation}
	\Omega_0(T, \mu, L, D) = k_BT \sum_{\mathbf{k}} \log\left(1 -e^{\beta \mu-\beta \varepsilon_{\mathbf{k}}(L,D)} \right),
\end{equation}     
where the summation is performed over $d$-dimensional wave vectors defined by Eq.~(\ref{spectrum}).\\ 
In the limit of infinite sidelength $L 
\to \infty$, the summation over the first $d-1$ variables may be replaced by integration. This yields
\begin{equation}
	\lim_{L \to \infty}\frac{1}{L^{d-1}} \Omega_0(T, \mu, L, D) =  -\frac{k_BT}{\lambda_t^{d-1}} \sum_{n_d \in \mathbb{Z}}g_{\frac{d + 1}{2}}\left(e^{\beta \mu - \beta \left(\tau \left(\frac{2 \pi n_d}{D}\right)^2 + \tau' \left(\frac{2 \pi n_d}{D}\right)^4 \right)}\right) \;, 
\label{Omega1} 
\end{equation}
where $\lambda_t$ (see below) is the thermal de Broglie length and $g_{\nu}(x) = \sum_{r=1}^{\infty} \frac{x^r}{r^{\nu}}$ denotes the Bose function. 
For later convenience, in addition to $\lambda_t$, we also introduce two other thermal lengthscales present in the system, related to the couplings $\tau$ and $\tau'$, such that:
\begin{equation}
\lambda_t = (4 \pi\beta t)^{1/2}\;,\;\;\; \lambda_\tau = (4 \pi\beta \tau)^{1/2}\;,\;\;\; \lambda_{\tau'}=\frac{4\pi}{\Gamma(1/4)}(\beta\tau')^{1/4}\;, 
\end{equation}
where $\beta^{-1}=k_B T$, $\Gamma(x)$ is the Euler gamma function, and we conveniently absorbed some numerical factors into the definitions of the thermal lengthscales. Throughout the analysis, we consider $D$ much larger than each of these scales (for an analysis of the opposite situation with standard dispersion see Ref.~\onlinecite{Jakubczyk_2016_2}), but make no assumptions concerning the ratios between $\lambda_t$, $\lambda_{\tau}$, and $\lambda_{\tau'}$.  \\
We now apply the Poisson formula,  conveniently rewritten in the form
\begin{equation}
	   		\sum_{n \in \mathbb{Z}}\psi(n) = \int_{\mathbb{R}}dx \psi(x) + 2\sum_{p=1}^{\infty}\int_{\mathbb{R}}dx \psi(x) \cos\left(2\pi p x\right) 
\end{equation}
to Eq.~(\ref{Omega1}). The first term [coming from $\int_{\mathbb{R}}dx  \psi(x)$] recovers $D\omega_b(T,\mu)$, the latter one is identified as $\omega_s(T,\mu,D)$ by virtue of Eq.~(\ref{decomposition}). By a change of integration variable and replacement of the summation with an integral using the Cauchy-Maclaurin formula (for a detailed justification of this step see Ref.~\onlinecite{Lebek_2021}), we obtain the following expression: 
\begin{equation}
   			\omega_{0, s}(T, \mu, D) =  -\frac{8k_BT}{\pi^{\frac{d+1}{2}} D^{d-1}}\frac{\lambda_{\tau}^{d-1}}{ \lambda_t^{d-1}}\sum_{p=1}^{\infty}\frac{1}{(2p)^d} \int_0^{\infty}d \eta \eta^{d-1}e^{\beta \mu \frac{p^2 D^2}{\eta^2 \beta \tau}}\int_0^{\infty}dx e^{-\left(x^2 + \frac{\gamma}{p^2}\eta^2x^4\right)}\cos(\eta x)\;. 
      \label{omega1}
      		\end{equation} 
Above we introduced 
\begin{equation}
\gamma = \frac{\tau'}{\tau D^2} = \frac{\Gamma(1/4)^4}{(4\pi)^3}\frac{{\lambda_{\tau'}}^4}{\lambda_{\tau}^2 D^2}  \;,     
\end{equation}
which will shortly be identified as a (dimensionless) scaling variable. 
The complexity of the present problem as compared to the previously studied cases of purely quadratic and purely quartic dispersion relations is evident from Eq.~(\ref{omega1}). If we take $\gamma=0$, the integrals are performed analytically, recovering previously known results \cite{Martin_2006}. On the other hand, by dropping $x^2$ in the last exponential of Eq.~(\ref{omega1}), we recover the case of purely quartic dispersion studied in Refs \cite{Lebek_2020, Lebek_2021, Napiorkowski_2021}. It clearly follows from the general expression of Eq.~(\ref{omega1}) that the crossover between these two limiting situations is governed by the parameter $\gamma$, which may be controlled by manipulating  either $\frac{\tau'}{\tau}$ or $D$. The major goal of the present study is to understand how the crossover between the two completely distinct situations (characterized in particular by different Casimir force signs in $d=3$) occurs without resorting to any expansion corresponding to $\gamma$ being small or large. 

With this goal in mind, by a sequence of transformations described in the Appendix, we now recast Eq.~(\ref{omega1}) in the following form: 
\begin{equation}
   			\beta \omega_{0, s}(T, \mu, D) = - \frac{\lambda_{\tau}^{d-1}}{\lambda_{t}^{d-1}}\frac{\Delta_0(x, \gamma, d)}{D^{d-1}}\;, 
\label{omega0s}      
\end{equation}
   		where
\begin{multline}
   			\Delta_0(x, \gamma, d) = \\  \frac{4}{(2 \pi)^{\frac{d}{2}}} \sum_{p=1}^{\infty}\sum_{k=0}^{\infty}\frac{(-8 \gamma)^k}{k!}\left(\frac{1}{4}\right)_k \left(\frac{3}{4}\right)_k\sum_{m=0}^{2k}\frac{1}{m!}\frac{\left(-2k \right)_m }{2^m\left(\frac{1}{2} \right)_m}x^{\frac{d}{2} + k + m} \frac{K_{
   					\frac{d}{2} + k + m}(px)}{p^{\frac{d}{2} + k - m}}\;. 
        \label{Delta0}
\end{multline}
Here $\left(n \right)_k$ denotes the Pochhammer symbol, $K_\alpha (z)$ is the hyperbolic Bessel function,  while 
\begin{equation}
x= D\sqrt{\frac{-\mu}{\tau}} = \frac{D}{\xi_2}   
\end{equation}
is a scaling variable recognized as the ratio between $D$ and the correlation length $\xi_2=\sqrt{\frac{\tau}{|\mu|}}$, measured along the direction perpendicular to the confining walls in the case of purely quadratic dispersion ($\tau' = 0$). The quantity $\xi_2$ diverges at the critical point, where $\mu=0$. Also note that $\xi_2$  measures the asymptotic (long-distance) decay of the correlation function in the $d$-th direction provided $\tau>0$. Eq.~(\ref{omega0s}) and Eq.~(\ref{Delta0}) hold under this assumption and the scaling variable $x$ may take arbitrary positive values. 

Eq.~(\ref{omega0s}) suggests that the Casimir energy decays as $1/D^{d-1}$ and the amplitude is governed by a universal scaling function $\Delta_0(x, \gamma, d)$ only after factoring out the nonuniversal coefficient $\frac{\lambda_{\tau}^{d-1}}{\lambda_{t}^{d-1}}$, which vanishes if $t=\tau$. This result is fully in line with expectations stemming from studies of other anisotropic systems. We finally point out that, upon putting $\gamma=0$, we recover the known result  
\begin{equation}
\Delta_0(x, \gamma =0, d) = \frac{4}{(2 \pi)^{\frac{d}{2}}} \sum_{p=1}^{\infty}x^{\frac{d}{2}} \frac{K_{
   					\frac{d}{2}}(px)}{p^{\frac{d}{2}}}
\end{equation}
obtained for the purely quadratic dispersion \cite{Napiorkowski_2022}. This result is in agreement with the $x \gg 1$ asymptotics of the scaling function in the $O(N)$ model in the Gaussian approximation obtained in previous works \cite{krech_diet_1992}. The expressions of Eq.~(\ref{omega0s}) and (\ref{Delta0}) are to our knowledge new, and are valid for all values of ($x$, $\gamma$). We point out that the standard scaling limit considered in ample literature on Casimir interactions involves sending $D$ to infinity and at the same time considering $T$ increasingly close to $T_c$, such that the dimensionless quantity $D/\xi$, where $\xi$ is the bulk correlation length, may take an arbitrary value. The key present generalization amounts to allowing for an additional dispersion parameter $\tau'$, which gives rise to the emergence of the new scaling variable $\gamma$.    

In what follows, we analyze the formula for the scaling function $\Delta_0(x, \gamma, d)$ at finite $\gamma$ with particular focus on the critical state ($x\to 0$) and odd integer spatial dimensionalities, where, as we demonstrate, it becomes surprisingly simple and meaningful.   

\subsection{Critical Casimir amplitude}
Upon putting $x=0$, by a sequence of exact transformations described in the Appendix, Eq.~(\ref{Delta0}) can be recast in the following form: 
\begin{equation}
   			\Delta_0(x=0, \gamma, d) = \frac{2}{\pi^{\frac{d}{2}}}\Gamma\left(\frac{d}{2}\right)\sum_{k=0}^{\infty}\frac{(-16 \gamma)^k}{k!}\left( \frac{1}{4}\right)_k \left( \frac{3}{4}\right)_k\left( \frac{d}{2}\right)_k \zeta(d + 2k) {}_2F_1\left(-2k, \frac{d}{2} + k; \frac{1}{2}; 1\right)\;, 
      \label{Delta01} 
\end{equation} 
where the generalized hypergeometric function of type $_k F_m$ is defined by the series 
\begin{equation}
	   		{}_kF_m\left(a_1, a_2, \ldots, a_k; c_1, c_2, \ldots, c_m;z\right) := \sum_{n=0}^{\infty}\frac{\left(a_1\right)_n\left(a_2\right)_n\ldots\left(a_k\right)_n}{\left(c_1\right)_n\left(c_2\right)_n\ldots\left(c_m\right)_n}\frac{z^n}{n!}\;.  
      \label{hyper}
\end{equation}
Here $z \in \mathbb{C}$, $a_1, a_2, \ldots, a_k \in \mathbb{C}$, and  $c_1, c_2, \ldots, c_m \in \mathbb{C} \backslash \left\{0, -1, -2, \ldots\right\}\; $. 

The key present observation, demonstrated in the Appendix, is that for $d=2l+1$, $l\in \mathbb{Z}_+$, the series of Eq.~(\ref{Delta01}) truncates such that $\Delta_0(x=0, \gamma, d=2l+1)$ turns out to be a polynomial of degree $l$ in the scaling variable $\gamma$. Strikingly, for the particular case of $d=3$, one obtains that $\Delta_0(x=0, \gamma, d=3)$ is a polynomial of degree 1 and in fact could be reconstructed from the known asymptotic expressions at vanishing $\tau$ (purely quartic dispersion in the $d$-th direction) and at vanishing $\tau'$ (purely quadratic dispersion in the $d$-th direction).  

\begin{figure*}
\includegraphics{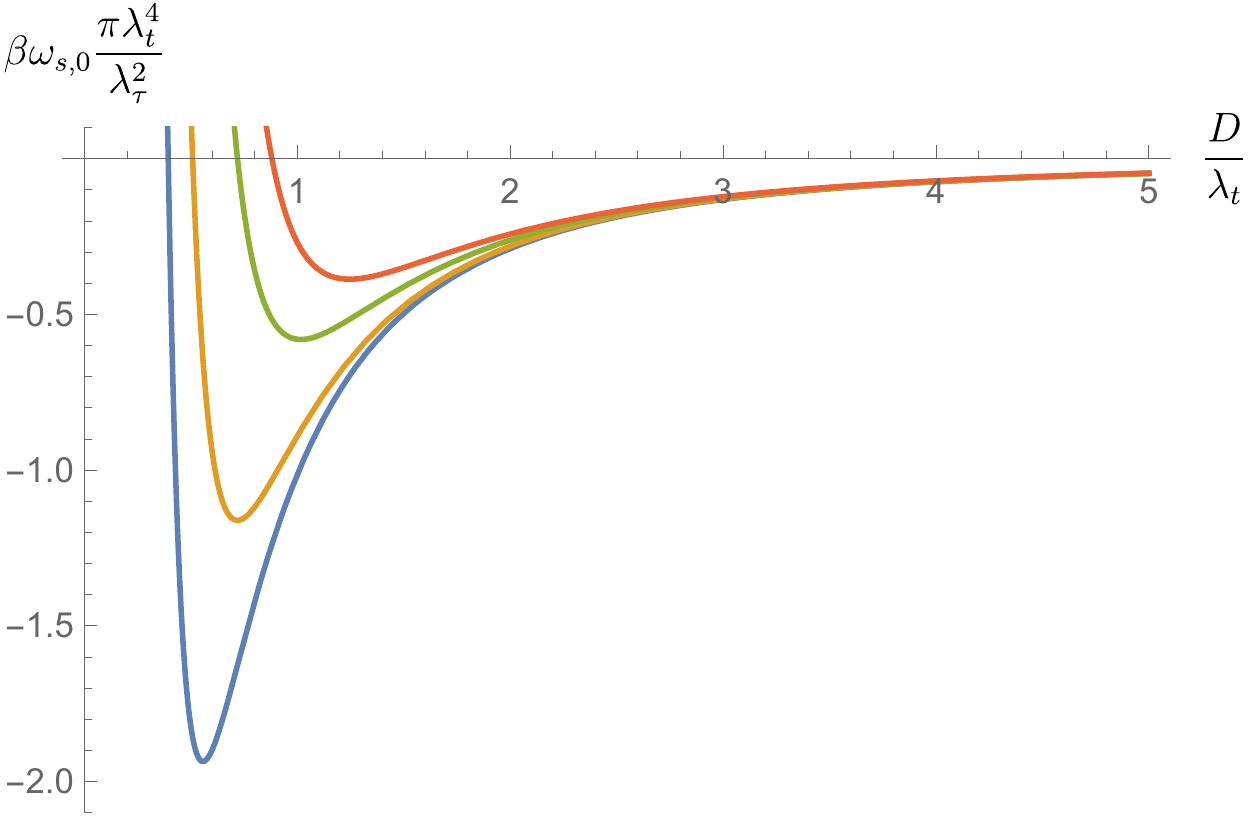}
\caption{\label{fig:wide} The dimensionless surface grand canonical potential for $d = 3$. The curves correspond to  $\frac{\lambda_{\tau'}^4}{\lambda_{\tau}^2 \lambda_t^2} \frac{\Gamma(1/4)^4}{(4\pi)^3} \in \{0.015, 0.025, 0.05, 0.075 \}$,  chosen this way for the clarity of the plot. The quantity $\omega_{s,0}$ acts as an effective potential binding the macroscopic objects  (``walls'') immersed in the fluid. Their equilibrium separation ($D^*$) corresponds to the minimum of $\omega_{s,0}$ and diverges for $\tau\to 0$ ($\lambda_\tau\to\infty$).} 
\label{Figd3}
\end{figure*}

\begin{figure*}
\includegraphics{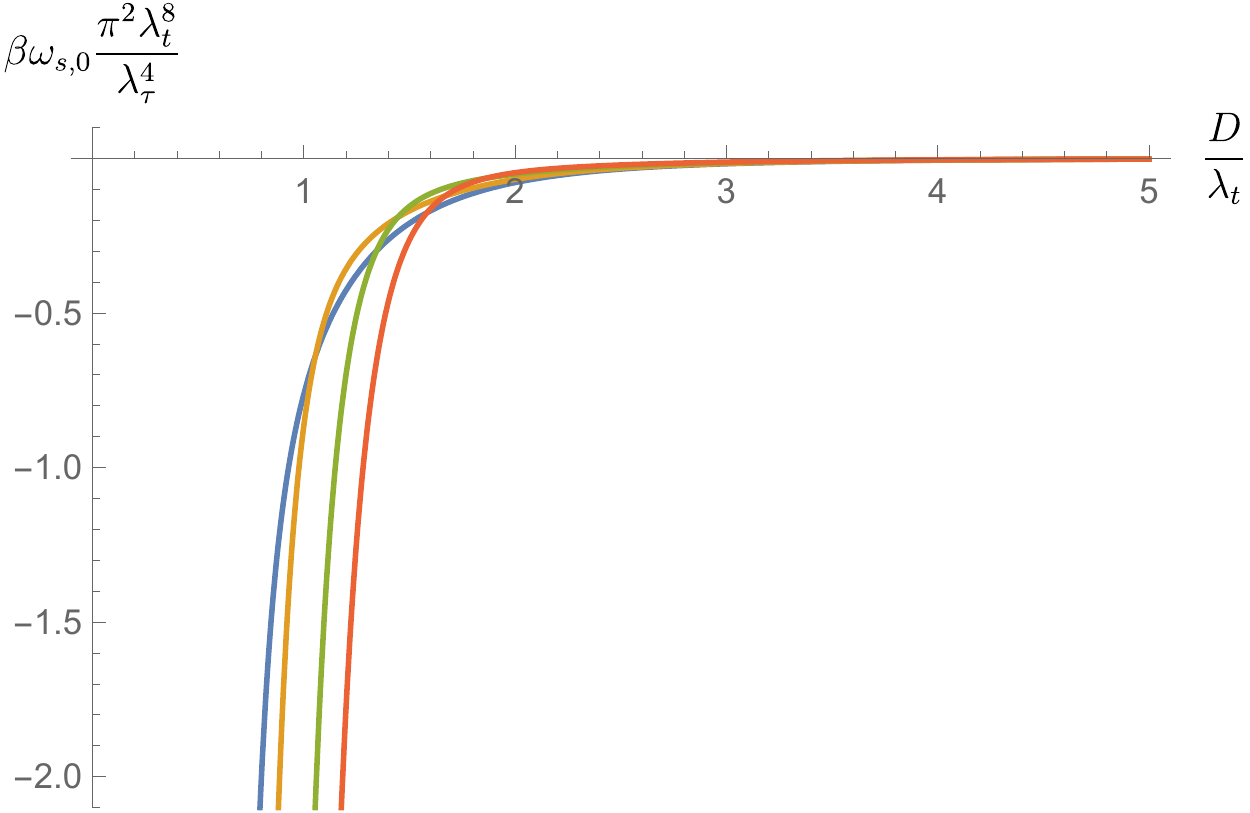}
\caption{\label{fig:wide} The dimensionless surface grand canonical potential for $d = 5$. The plot corresponds to $\frac{\lambda_{\tau'}^4}{\lambda_{\tau}^2 \lambda_t^2} \frac{\Gamma(1/4)^4}{(4\pi)^3}  \in \{0.015, 0.025, 0.05, 0.075 \}$.  The function is monotonously increasing and the Casimir force is attractive for all values of $\frac{D}{\lambda_t}$ and all choices of the system parameters. There is no possibility of obtaining a bound state characterized by a minimum of $\omega_{s,0}$ at finite $D$. } 
\label{Figd5}
\end{figure*}

Specifically, for $d=3$, we obtain: 
\begin{equation}
   	 	\omega_{0, s}(T, 0, D) = -\frac{ \zeta(3)}{\pi}\frac{k_BT}{D^2}\frac{\tau}{t} + \frac{12\zeta(5)}{\pi}\frac{k_BT}{D^4}\frac{\tau'}{t}\;. 
\label{omegad3}
\end{equation}
As already indicated, the resulting formula is simply the sum of expressions for the surface contributions to the grand canonical free energy in a system with purely quadratic dispersion (the negative term) and with purely quartic dispersion (the positive term). This expression is of course in agreement with the previously obtained results, but is far from trivial and its simplicity appears striking to us. The Casimir potential as obtained in Eq.~(\ref{omegad3}) is plotted in Fig.~\ref{Figd3}. It is governed by the negative (attractive) contribution at asymptotically large separation $D$ and a positive (repulsive) one at $D$ smaller. 
For comparison, we also give explicit expression for $d=5$ (corresponding to $l=2$, where $\Delta_0(x=0, \gamma, d=5)$ is a polynomial of degree $l=2$ in $\gamma$). For this case we find:   
 \begin{equation}
   	 	\omega_{0, s}(T, 0, D) = -\frac{3 \zeta(5)}{2\pi^2}\frac{k_BT}{D^4}\left(\frac{\tau}{t}\right)^2 + \frac{90 \zeta(7)}{\pi^2}\frac{k_BT}{D^6}\frac{\tau \tau'}{t^2} - \frac{2520 \zeta(9)}{\pi^2 }\frac{k_BT}{D^8}\left(\frac{\tau'}{t}\right)^2\;, 
\label{omegad5}
\end{equation} 
which is plotted in Fig.~\ref{Figd5}. 
Note that in Eq.~(\ref{omegad5}) [unlike Eq.~(\ref{omegad3})], both the terms governing the short- ($\sim1/D^8$) and long- ($\sim1/D^4$) distance behavior carry the negative sign. This is reflected in the generically attractive character of the Casimir interaction and absence of a bound state characterized by a finite equilibrium separation $D$ for any values of the system parameters.    

\begin{figure*}
\includegraphics{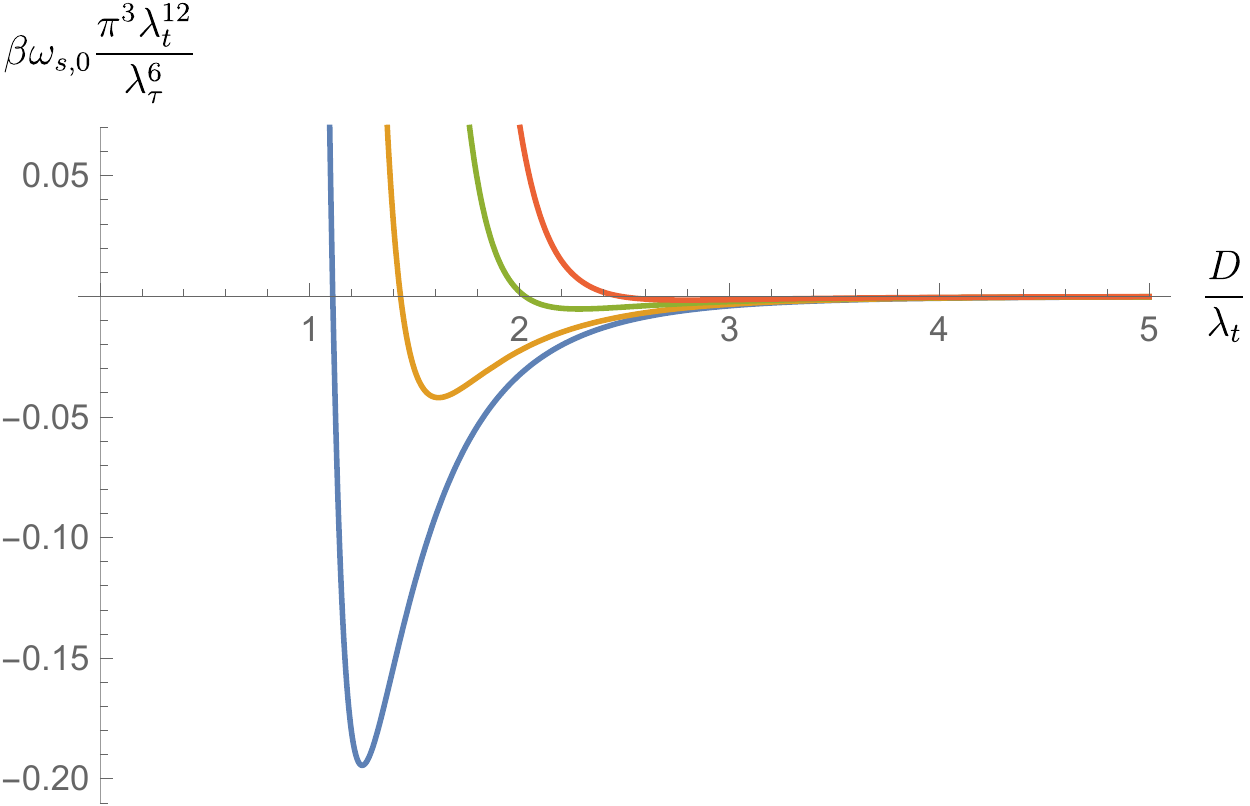}% Here is how to import EPS art
\caption{\label{fig:wide} The dimensionless surface grand canonical potential for $d = 7$. The plot is sketched for $\frac{\lambda_{\tau'}^4}{\lambda_{\tau}^2 \lambda_t^2} \frac{\Gamma(1/4)^4}{(4\pi)^3} \in \{0.015, 0.025, 0.05, 0.075 \}$. The surface contribution to the grand canonical potential features a unique minimum.} 
\label{Figd7}
\end{figure*}

For further comparison, we also plot $\omega_{0, s}(T, 0, D)$ for $d=7$ in Fig.~\ref{Figd7} . This qualitatively resembles the situation obtained in $d=3$. 
In contrast to the cases $d=3$ and $d=7$, the effective Casimir potential $\omega_{0, s}(T, 0, D)$ corresponding to $d=5$ features no minimum and appears attractive for all values of $D$. We verified numerically (up to $d=23$), that $\omega_{0, s}(T, 0, D)$ yields purely attractive interaction for $d\in\{5,9,13,\dots\}$. For $d\in\{3,7,11,\dots\}$, the profile of $\omega_{0, s}(T, 0, D)$ is qualitatively similar to the one obtained for $d=3$, i.e. it exhibits a single minimum corresponding to an equilibrium separation of the walls. From the numerical analysis we obtain that there is no possibility of generating a Casimir potential featuring multiple equilibrium configurations (stable or not) for the presently analyzed system, the number of extrema of $\omega_{s,0}$ at finite $D$ is either zero or one. We conclude that the two qualitatively different situations depicted in Figs.~\ref{Figd3} and \ref{Figd5} exhaust the spectrum of possibilities (at least up to $d=23$).   

We end this subsection by briefly commenting on the case of even dimensionalities $d\in\{4,6,8,\dots\}$, where we were not able to obtain simple analytical results. Eq.~(\ref{omega0s}) and Eq.~(\ref{Delta0}) hold also for this case, however the truncation of the series of Eq.~(\ref{Delta01}) analogous to the one exploited for $d\in\{3,5,7,\dots\}$ does not occur. From the previously studied behavior of the Casimir energy $\omega_s$ for purely quartic dispersion we know that the corresponding Casimir amplitude strictly vanishes \cite{Lebek_2021, Napiorkowski_2021} for $d\in\{4, 6, 8, \dots\}$. Since the Casimir force is attractive for purely quadratic dispersion, there is no obvious mechanism that might generate a repulsive term in $\omega_{s,0}$ and in consequence no bound state due to the Casimir interaction should, in our opinion, be expected.    

\subsection{Equilibrium separation}
Focusing on $d=3$, we note that the equilibrium separation $D^*$  between the walls, evaluated from $\frac{\partial \omega_{0, s}(T, 0, D)}{\partial D}|_{D=D^*}=0$, is given by: 
\begin{equation}
D^*=\sqrt{\frac{24\zeta(5)}{\zeta(3)}}\sqrt{\frac{\tau'}{\tau}}    
\end{equation}
and diverges for $\tau\to 0$ following the universal power law $D^*\sim \tau^{-1/2}$. This behavior, together with the shape of the effective binding potential $\omega_{0, s}(T, 0, D)$ is reminiscent of unbinding phenomena \cite{Dietrich_Domb, Forgacs_Domb} studied intensively e.g. in the contexts of fluid wetting or membrane unbinding. Also note the vanishing of the effective potential curvature  at the minimum 
\begin{equation}
\frac{\partial^2\omega_{0,s}}{\partial D^2}|_{D=D*}\sim\frac{\tau^3}{\tau'^2}\;,    
\end{equation}
indicating the increasing magnitude of fluctuations of $D^*$ for $\tau\to 0$. 

\subsection{Casimir interactions above $T_c$}
We now analyze thermodynamic states above the condensation temperature ($T>T_c$). We restrict to $d=3$ and note that for a system characterized by a purely quadratic dispersion, the expression for $\omega_{0,s}$ reads \cite{Gambassi_2006}: 
\begin{equation}
	\omega_{0,s}\left(T, \mu, D\right) = -\frac{k_BT}{\pi D^2}\left(\frac{\lambda_{\tau}}{ \lambda_t}\right)^2\left(g_3\left(e^{-\frac{D}{\xi_2}}\right) + \frac{D}{\xi} g_2\left(e^{-\frac{D}{\xi_2}}\right) \right)\;, 
 \label{omega3}
\end{equation}
which decays exponentially for $D\gg \xi_2$. The corresponding expression for $\tau=0$ and $\tau'>0$ is given by 
\begin{equation}
	\beta\lambda_t^{d-1}\omega_{0,s}\left(T, \mu, D\right) \sim \frac{1}{D^{\alpha}}\sum_{p=1}^{\infty}e^{-\frac{pD}{\xi_4}}\cos\left(\frac{pD}{\xi_4}\right) ,\qquad \textrm{for} \;\;\; \frac{D}{\lambda_{\tau'}} \gg 1\; 
 \label{omega4}
\end{equation}
Here $\xi_4 = \left(\frac{4\tau'}{|\mu|}\right)^{\frac{1}{4}}$ describes the correlation length in the direction along the $d$-th direction. Note that the behavior of $\omega_{0,s}$ is completely different in the two situations described by Eq.~(\ref{omega3}) and Eq.~(\ref{omega4}). In the latter case one obtains exponentially damped oscillations, so that $\omega_{0,s}$ periodically changes sign as a function of $D$, unlike the case described by Eq.~(\ref{omega3}), where the behavior is purely monotonous. These two distinct behaviors reflect the structure of the correlation function in the bulk anisotropic fluid as discussed in Ref.~\onlinecite{Jakubczyk_2018}. It is interesting to investigate the evolution of the picture in presence of both $\tau$ and $\tau'$. We focus on the exponential decay, leaving out the algebraic decay factor $D^\alpha$. For this aim, it is sufficient to consider the integral 
\begin{equation}
I_p = \int_{0}^{\infty} d \eta \int_{\mathbb{R}}d\tilde{x}e^{- \left(\beta|\mu|\eta + \beta \tau \eta \left(\frac{2\pi \tilde{x}}{D} \right)^2 + \beta \tau' \eta \left(\frac{2\pi \tilde{x}}{D} \right)^4 + 2\pi i p \tilde{x}\right)}\;.  
\label{saddle1}
\end{equation}
By the variable transformation $\tilde{x} \mapsto x = 2 \pi \frac{\xi_2}{D} u$ and $\eta \mapsto t = \frac{\xi_2}{D} \eta$, we rewrite $I_p$ in the form 
\begin{equation}
I_p = \frac{1}{2\pi}\left(\frac{D}{\xi_2}\right)^2\int_{0}^{\infty} dt \int_{\mathbb{R}}dx e^{-\frac{D}{\sqrt{\beta \tau}}\phi_p(t,x)} \;,
\label{saddle2}
\end{equation}
where 
\begin{equation}
	\phi_p\left(t,x\right) = b^3t + b^3 x^2 t + \frac{1}{4}a^4b^3 x^4 t + ipbx,
 \label{saddle3}
\end{equation} 
with $b = \sqrt{\beta |\mu|}$ and $a = \left(\frac{4\tau' |\mu|}{\tau^2}\right)^{\frac{1}{4}} = \left(\frac{\xi_{4}}{\xi_{2}}\right)$. 
The above expression for $I_p$ is clearly susceptible to the saddle-point analysis, which becomes exact for $D/\lambda_t\to\infty$. We postpone the details of the analysis to the Appendix and summarize our major findings below. Our results indicate that the saddle-point equations display qualitatively different solutions, depending on whether the parameter $a$ is larger or smaller than unity. For $a<1$ we find that $I_p$ decays as $e^{-D/\xi_>}$, where 
\begin{equation}
\frac{1}{\xi_>}=\frac{1}{\xi_2}   \left[1 + \frac{1}{2} \frac{\tau' |\mu|}{\tau^2} + \frac{7}{8} \left(\frac{\tau' |\mu|}{\tau^2}\right)^2 + O\left(\left(\frac{\tau' |\mu|}{\tau^2}\right)^3\right)\right]. 
\label{corr1}
\end{equation}
At leading order in $|\mu|$ (recall that $\mu=0$ corresponds to the condensation transition) one finds $\xi_>=\xi_2$. Corrections appear as a power series in $\frac{\tau' |\mu|}{\tau^2}=\frac{1}{4}a^4$. In particular, the divergence of $\xi_>$ at condensation is the same as for $\xi_2$. The situation appears very different for $a>1$. At the technical level (see the Appendix), this follows from the existence of two physically relevant saddle-point solutions, being mutual complex conjugate. In this case we find:
\begin{multline}
	I_p \sim e^{-\frac{D}{\sqrt{\beta \tau}}\phi_p\left(\tilde{t}^{(a)},\tilde{x}^{(a)}\right)} + e^{-\frac{D}{\sqrt{\beta \tau}}\phi_p\left(\tilde{t}^{(b)},\tilde{x}^{(b)}\right)}  = \\ 2e^{-Dp \left(\frac{|\mu|}{\tau'}\right)^{\frac{1}{4}}\cos\left(\frac{1}{2}\arctan{\sqrt{\frac{4\tau' |\mu|}{\tau^2}-1}}\right)}\cos{\left(Dp \left(\frac{|\mu|}{\tau'}\right)^{\frac{1}{4}}\sin{\left(\frac{1}{2}\arctan{\sqrt{\frac{4\tau' |\mu|}{\tau^2} -1}}\right)}\right)}\;.
 \label{ocs1}
\end{multline}
The above expression, when viewed as a function of $D$, features exponentially damped oscillations. This bears a resemblance to the asymptotic behavior of the surface contribution to the grand canonical potential in the system with purely quartic dispersion, which can be continuously recovered by going to the limit $\tau \to 0$. Indeed, as $\tau \to 0$, one has $a \to \infty$. This implies $\arctan\sqrt{a^4 - 1} \to \frac{\pi}{2}$ and
\begin{equation}
	I_p \sim e^{-\frac{D}{\sqrt{\beta \tau}}\phi_p\left(\tilde{t}^{(a)},\tilde{x}^{(a)}\right)} + e^{-\frac{D}{\sqrt{\beta \tau}}\phi_p\left(\tilde{t}^{(b)},\tilde{x}^{(b)}\right)} \xrightarrow[\tau \to 0]{} 2e^{-Dp\left(\frac{|\mu|}{4\tau'}\right)^{\frac{1}{4}}} \cos{\left(Dp \left(\frac{|\mu|}{4\tau'}\right)^{\frac{1}{4}}\right)}\;,
 \label{ocs2}
\end{equation}
which is exactly the behavior of the system with purely quartic dispersion in the $d$-th direction.   For $a \gg 1$\textbf{,} the dominant term of inverse of the bulk correlation length coincides with $\frac{1}{\xi_4}$ and the corrections are expressed via a series expansion in $\frac{\tau}{\sqrt{\tau' |\mu|}} = \frac{2}{a^2}$ 
\begin{equation}
	\frac{1}{\xi_<} = \frac{1}{\xi_4}\left(1 + \frac{1}{4} \frac{\tau}{\sqrt{\tau' |\mu|}} -\frac{1}{32} \frac{\tau^2}{\tau' |\mu|} + O \left(\left(\frac{\tau}{\sqrt{\tau' |\mu|}}\right)^3\right)\right). 
 \label{corr2}
\end{equation}
In summary, we have demonstrated that, at least as the leading behavior is concerned, the oscillatory behavior of $\omega_{0,s}$ appears once the parameter $a=\frac{\xi_4}{\xi_2}$ exceeds unity. The scaling of the correlation length with $|\mu|$ is governed by the exponent $\nu=\frac{1}{2}$ for $a<1$ and $\nu=\frac{1}{4}$ for $a>1$.

\section{Summary}
We have analyzed the Casimir effect in an ideal Bose gas characterized by a dispersion relation involving both terms quadratic and quartic in momentum. Our motivation was to continuously connect two previously studied cases of purely quadratic and purely quartic dispersions in the direction perpendicular to the confining walls and understand the corresponding crossover behavior. These two asymptotic cases display, in physically most relevant ranges of spatial dimensionality $d$, very distinct behavior characterized in particular by different exponents controlling the power-law decay of the Casimir intertaction, as well as opposite Casimir force sign. We have derived a general, rather complicated expression for the Casimir energy, which allowed us to identify the two relevant scaling variables. Restricting to thermodynamic states at Bose-Einstein condensation, we have demonstrated existence of a single minimum of the Casimir energy viewed as a function of the separation $D$ between the confining walls for dimensionalities $d\in\{3,7,11,\dots\}$ and the lack of such a minimum for $d\in\{5,9,13,\dots\}$. In these cases, the Casimir energy can be expressed as a polynomial in the identified scaling variable.  The obtained minima describe stable equilibrium position of the confining walls. Subsequently, we departed from $T=T_c$ and investigated the interfacial contribution to the grand canonical potential at $T>T_c$. For $d=3$, we obtained either monotonous exponential decay or exponential decay modulated by periodic oscillations, depending on the parameter $a=\frac{\xi_4}{\xi_2}$ expressing the ratio between bulk correlation lengths in the systems with purely quadratic and purely quartic dispersion relations in the direction perpendicular to the walls.   

Our analysis is restricted to the non-interacting Bose gas. We point out however that, in the previously studied cases involving either quadratic or quartic dispersions, the ideal and mean-field (imperfect)  Bose gases exhibit very similar behavior, at least at and above $T_c$. It is therefore natural to expect an analogous situation in the presently studied case. The imperfect Bose gas in turn is closely related to the $N\to\infty$ limit of the $O(N)$-symmetric models. We therefore anticipate that our results should also be pertinent to the $O(N)$ class models for $N$ large enough.  

\begin{acknowledgments}
We are grateful to Jacek Jezierski for the help with properly interpreting the mathematical objects that emerged in the calculations. We thank Marek Napi\'orkowski for discussions as well as comments on the manuscript. PJ acknowledges support from the Polish National Science Center via grant 2021/43/B/ST3/01223.
\end{acknowledgments}

\appendix 
\section{Grand Canonical Potential}
	 \subsection{Inductive derivation of $\int_{\mathbb{R}}dxe^{-(x^2 + 2 i px)}x^{2k}$}
  In order to obtain the expressions for the surface contribution to the grand canonical potential Eq.~(\ref{omega0s}) and Eq.~(\ref{Delta0}), one first needs to prove  that
		\begin{equation}
			\int_{\mathbb{R}}dxe^{-(x^2 + 2 i px)}x^{2k} = \sqrt{\pi}e^{-p^2}\frac{(2k-1)!!}{2^k}{}_1F_1\left(-k; \frac{1}{2}; p^2 \right)
			, \qquad \forall k\in \mathbb{Z}_+.
    \label{ind1}
		\end{equation}
	We present the proof of this lemma. \\   We notice that 
	\begin{equation*}
		\int_{\mathbb{R}}dxe^{-(x^2 + 2 i px)}x^{2k} = \lim_{a \to 1} (-1)^k \left(\frac{\partial}{\partial a}\right)^k\int_{\mathbb{R}}dxe^{-(ax^2 + 2 i px)} = \lim_{a \to 1} (-1)^k \left(\frac{\partial}{\partial a}\right)^k \sqrt{\frac{\pi}{a}}e^{-\frac{p^2}{a}}.
	\end{equation*}
	To prove Eq.~(\ref{ind1}), we prove an even  stronger statement
	\begin{equation} \left(\frac{\partial}{\partial a}\right)^k \sqrt{\frac{\pi}{a}}e^{-\frac{p^2}{a}} = \sqrt{\frac{\pi}{a^{2k + 1}}}e^{-\frac{p^2}{a}}\frac{(2k-1)!!}{2^k}(-1)^k{}_1F_1\left(-k; \frac{1}{2}; \frac{p^2}{a} \right)
 \label{ind2}
	\end{equation}
	using the principle of mathematical induction. \\ One easily sees that for $k=1$, the identity Eq.~(\ref{ind2}) is satisfied. Our goal is to resolve whether the fact that the identity holds for some arbitrary $k$ automatically implies that it also holds for $k+1$. Let us assume that  $\exists k \in \mathbb{Z}_+$ such that Eq.~(\ref{ind2}) is true (for that $k$). \\ Then
	\begin{multline*}  \left(\frac{\partial}{\partial a}\right)^{k+1}\left[ \sqrt{\frac{\pi}{a}}e^{-\frac{p^2}{a}}\right] = \left(\frac{\partial}{\partial a}\right)\left\{\left(\frac{\partial}{\partial a}\right)^{k}\left[ \sqrt{\frac{\pi}{a}}e^{-\frac{p^2}{a}}\right]\right\} = \\ \left(\frac{\partial}{\partial a}\right)\left\{ \sqrt{\frac{\pi}{a^{2k + 1}}}e^{-\frac{p^2}{a}}\frac{(2k-1)!!}{2^k}(-1)^k{}_1F_1\left(-k; \frac{1}{2}; \frac{p^2}{a} \right)\right\} = \\   \sqrt{\frac{\pi}{a^{2(k+1) + 1}}}e^{-\frac{p^2}{a}}\frac{(2(k+1)-1)!!}{2^{k+1}}(-1)^{k+1} \times \\ \left\{{}_1F_1\left(-k; \frac{1}{2}; \frac{p^2}{a} \right) - \frac{p^2}{a}\frac{2}{2k+1}{}_1F_1\left(-k; \frac{1}{2}; \frac{p^2}{a} \right) - \frac{2}{2k + 1}a \frac{\partial}{\partial a}\left[{}_1F_1\left(-k; \frac{1}{2}; \frac{p^2}{a} \right)\right]\right\}.
	\end{multline*}
	What remains to be proven is if 
	\begin{equation}
		\left[1 -\frac{2}{2k+1}\frac{p^2}{a} - \frac{2}{2k + 1}a \frac{\partial}{\partial a}\right]
		{}_1F_1\left(-k; \frac{1}{2}; \frac{p^2}{a} \right) = {}_1F_1\left(-(k+1); \frac{1}{2}; \frac{p^2}{a} \right).
  \label{ind3}
	\end{equation} 
 	We have
 	\begin{equation*}
 		{}_1F_1\left(-k; \frac{1}{2}; \frac{p^2}{a} \right)= \sum_{m=0}^k (-1)^m \binom{k}{m}\frac{2^m}{(2m-1)!!}\left(\frac{p^2}{a}\right)^m,
 	\end{equation*}
 	and
	\begin{equation*}
		-\frac{2}{2k+1}\frac{p^2}{a}{}_1F_1\left(-k; \frac{1}{2}; \frac{p^2}{a} \right)= \frac{1}{2k +1}\sum_{m=0}^k (-1)^{m+1} \binom{k}{m}\frac{2^{m+1}}{(2m-1)!!}\left(\frac{p^2}{a}\right)^{m+1},
	\end{equation*}
	and
	\begin{equation*}
		-\frac{2}{2k+1}a \frac{\partial}{\partial a}{}_1F_1\left(-k; \frac{1}{2}; \frac{p^2}{a} \right)= \frac{1}{2k +1}\sum_{m=0}^k (-1)^{m}m \binom{k}{m}\frac{2^{m+1}}{(2m-1)!!}\left(\frac{p^2}{a}\right)^{m+1}.
	\end{equation*}
	After simple manipulations on the indices of summation, the left hand side of Eq.~(\ref{ind3}) may be rewritten as
	\begin{equation*}
		1 + \sum_{m=1}^k\frac{(-1)^m2^m}{(2m-1)!!}\left(\frac{p^2}{a}\right)^m\left[\binom{k}{m} +\frac{2m -1}{2k + 1}\binom{k}{m-1} + \frac{2m}{2k + 1}\binom{k}{m}\right] + \frac{(-1)^{k+1}2^{k+1}}{(2(k+1)-1)!!}\left(\frac{p^2}{a}\right)^{k+1}.
	\end{equation*}
	After expanding the right hand side of Eq.~(\ref{ind3}), one sees that the induction step is correct if for $m \in \{1,2, \ldots, k\}$
	\begin{equation*}
		\binom{k}{m} +\frac{2m -1}{2k + 1}\binom{k}{m-1} + \frac{2m}{2k + 1}\binom{k}{m} = \binom{k+1}{m}
	\end{equation*}
	is true. It indeed is, which is proven by simple algebraic manipulations. \\ 
	These considerations conclude the proof of Eq.~(\ref{ind1}).
		\subsection{Case of $\mu < 0$}
		Having presented the proof of the necessary lemma Eq.~(\ref{ind1}), we sketch the derivation of the expressions for the grand canonical potential Eq.~(\ref{omega0s}) and Eq.~(\ref{Delta0}). We start by casting Eq.~(\ref{Omega1}) in the form
		\begin{equation*}
		\beta \omega_{0,s}(T, \mu, D ) =-\frac{2}{\lambda_t^{d-1}} F_d(T, \mu, D),
		\end{equation*}
		where	
		\begin{multline*}
			 F_d(T, \mu, D):=\sum_{p=1}^{\infty}\sum_{r=1}^{\infty}\int_{\mathbb{R}}\frac{du}{r^{\frac{d+1}{2}}}e^{- \left(\beta|\mu|r + \beta \tau r \left(\frac{2\pi u}{D} \right)^2 + \beta \tau' r \left(\frac{2\pi u}{D} \right)^4 + 2\pi i p u\right)} =  \\ \frac{1}{2\pi} \frac{D}{\sqrt{\beta \tau}} \sum_{p=1}^{\infty}\sum_{r=1}^{\infty}\frac{e^{-r\frac{\beta \tau}{\xi_2^2}}}{r^{\frac{d}{2} +1}}\int_{\mathbb{R}}dxe^{- \left(x^2 +  \frac{\gamma D^2}{\beta \tau r}x^4 +  ip x \frac{D}{\sqrt{\beta \tau r}}\right)},
		\end{multline*}		
		with $\gamma = \frac{\tau'}{\tau D^2}$ and $\xi_2 = \sqrt{\frac{\tau}{|\mu|}}$.  \\ We now expand in $\gamma$
		\begin{equation*}
			e^{- \frac{\gamma D^2}{\beta \tau r}x^4} = \sum_{k=0}^{\infty}\frac{1}{k!}\left( - \frac{\gamma D^2}{\beta \tau r}\right)^kx^{4k} 
		\end{equation*}
		and change the order of the integration and the summation
		\begin{equation*}
			F_d(T, \mu, D) = \frac{1}{2\pi} \frac{D}{\sqrt{\beta \tau}} \sum_{p=1}^{\infty}\sum_{k=0}^{\infty}\sum_{r=1}^{\infty}\frac{e^{-r\frac{\beta \tau}{\xi_2^2}}}{r^{\frac{d}{2} +k+1}}\left( - \frac{\gamma D^2}{\beta \tau }\right)^k\frac{1}{k!}\int_{\mathbb{R}}dxe^{- \left(x^2  +  ip x \frac{D}{\sqrt{\beta \tau r}}\right)}x^{4k}.
		\end{equation*} 
		We use the identity of Eq.~(\ref{ind1}) to obtain 
		\begin{equation*}
			\int_{\mathbb{R}}dxe^{- \left(x^2  +  ip x \frac{D}{\sqrt{\beta \tau r}}\right)}x^{4k} = \sqrt{\pi}e^{-\frac{p^2D^2}{4 \beta \tau r}}\frac{(4k-1)!!}{4^k}{}_1F_1\left(-2k; \frac{1}{2}; \frac{p^2D^2}{4 \beta \tau r} \right).
		\end{equation*}
		The algebraic identity  $\frac{(4k-1)!!}{4^k} = 4^k\left(\frac{1}{4}\right)_k \left(\frac{3}{4}\right)_k$  allows us to write
		\begin{multline*}
			F_d(T, \mu, D) =  \frac{D}{\lambda_{\tau}} \sum_{p=1}^{\infty}\sum_{k=0}^{\infty}\left( - \frac{4\gamma D^2}{\beta \tau }\right)^k\frac{1}{k!}\left(\frac{1}{4}\right)_k \left(\frac{3}{4}\right)_k \\ \times \sum_{r=1}^{\infty}\frac{e^{-\left(r\frac{\beta \tau}{\xi_2^2} + \frac{1}{r}\frac{p^2D^2}{4 \beta \tau}\right)}}{r^{\frac{d}{2} +k+1}}{}_1F_1\left(-2k; \frac{1}{2}; \frac{p^2D^2}{4 \beta \tau r} \right).
		\end{multline*} 
		We now expand the hypergeometric function and obtain
		\begin{multline*}
			F_d(T, \mu, D) =  \frac{D}{\lambda_{\tau}} \sum_{p=1}^{\infty}\sum_{k=0}^{\infty}\left( - \frac{4\gamma D^2}{\beta \tau }\right)^k\frac{1}{k!}\left(\frac{1}{4}\right)_k \left(\frac{3}{4}\right)_k \sum_{m=0}^{2k}\frac{1}{m!}\frac{\left( -2k\right)_m }{\left( \frac{1}{2}\right)_m}\left(\frac{p^2D^2}{4 \beta \tau}\right)^m \\ \times \sum_{r=1}^{\infty}\frac{e^{-\left(r\frac{\beta \tau}{\xi_2^2} + \frac{1}{r}\frac{p^2D^2}{4 \beta \tau}\right)}}{r^{\frac{d}{2} +k+m+1}}.
		\end{multline*} 
		We now notice that a function of the form $e^{-(r+ a/r)}/r^{\alpha}$ ($\alpha>0, a>0$) is integrable on $\mathbb{R}_+$ and all of its derivatives vanish at $r \to 0$ and $r \to \infty$. This, due to the Euler-Maclaurin formula (see Ref.~\onlinecite{Lebek_2021}), allows us to approximate the $r$-summation with an integration
		\begin{multline*}
			\sum_{r=1}^{\infty}\frac{e^{-\left(r\frac{\beta \tau}{\xi_2^2} + \frac{1}{r}\frac{p^2D^2}{4 \beta \tau}\right)}}{r^{\frac{d}{2} +k+m+1}} \approx \int_0^{\infty} \frac{d\eta }{\eta^{\frac{d}{2} +k+m+1}}e^{-\left(\eta\frac{\beta \tau}{\xi_2^2} + \frac{1}{\eta}\frac{p^2D^2}{4 \beta \tau}\right)} = \\  \left( \frac{\beta \tau}{\xi_2^2}\right)^{\frac{d}{2} + k + m} \int_0^{\infty}\frac{d\eta}{\eta^{\frac{d}{2} +k+m+1}} e^{-\left(\eta +  \frac{1}{\eta}\left(\frac{pD}{2\xi_2}\right)^2\right)} =  \\ 2^{ \frac{d}{2} + k + m + 1} \left( \frac{\beta \tau}{\xi_2^2}\right)^{\frac{d}{2} + k + m}\left(\frac{pD}{\xi_2}\right)^{-\left(\frac{d}{2} + k + m\right)}K_{\frac{d}{2}+k + m}\left(\frac{pD}{\xi_2}\right),
		\end{multline*}
		  where we introduced the scaling variable $x:= \frac{D}{\xi_2}$. \\  After some algebra, we arrive at the result
		\begin{equation*}
			F_d(T, \mu, D) = \frac{2}{(2 \pi)^{\frac{d}{2}}} \frac{\lambda_{\tau}^{d-1}}{D^{d-1}} \sum_{p=1}^{\infty}\sum_{k=0}^{\infty}\frac{(-8 \gamma)^k}{k!}\left(\frac{1}{4}\right)_k \left(\frac{3}{4}\right)_k\sum_{m=0}^{2k}\frac{1}{m!}\frac{\left(-2k \right)_m }{2^m\left(\frac{1}{2} \right)_m}x^{\frac{d}{2} + k + m} \frac{K_{
			\frac{d}{2} + k + m}(px)}{p^{\frac{d}{2} + k - m}},
		\end{equation*}
		which allows us to write
		\begin{equation*}
			\beta \omega_{0, s}(T, \mu, D) = - \frac{\lambda_{\tau}^{d-1}}{\lambda_{t}^{d-1}}\frac{\Delta_0(x, \gamma, d)}{D^{d-1}},
		\end{equation*}
		with 
		\begin{equation*}
			\Delta_0(x, \gamma, d) = \frac{4}{(2 \pi)^{\frac{d}{2}}} \sum_{p=1}^{\infty}\sum_{k=0}^{\infty}\frac{(-8 \gamma)^k}{k!}\left(\frac{1}{4}\right)_k \left(\frac{3}{4}\right)_k\sum_{m=0}^{2k}\frac{1}{m!}\frac{\left(-2k \right)_m }{2^m\left(\frac{1}{2} \right)_m}x^{\frac{d}{2} + k + m} \frac{K_{
					\frac{d}{2} + k + m}(px)}{p^{\frac{d}{2} + k - m}}.
		\end{equation*}
        This concludes the sketch of derivation of Eq.~(\ref{omega0s}) and Eq.~(\ref{Delta0}).
	\subsection{The critical point - limiting procedure}
        Having derived the expressions for the surface contribution to the grand canonical potential for $x>0$ (Eq.~(\ref{omega0s}) and Eq.~(\ref{Delta0})), we present the limiting procedure allowing for deriving Eq.~(\ref{Delta01}). We recall Eq.~(\ref{omega0s})
				\begin{equation*}
			\beta \omega_{0, s}(T, \mu, D) = - \frac{\lambda_{\tau}^{d-1}}{\lambda_{t}^{d-1}}\frac{\Delta_0(x, \gamma, d)}{D^{d-1}}
		\end{equation*}
	   and Eq.~(\ref{Delta0}) 
		\begin{equation*}
			\Delta_0(x, \gamma, d) = \frac{4}{(2 \pi)^{\frac{d}{2}}} \sum_{p=1}^{\infty}\sum_{k=0}^{\infty}\frac{(-8 \gamma)^k}{k!}\left(\frac{1}{4}\right)_k \left(\frac{3}{4}\right)_k\sum_{m=0}^{2k}\frac{1}{m!}\frac{\left(-2k \right)_m }{2^m\left(\frac{1}{2} \right)_m}x^{\frac{d}{2} + k + m} \frac{K_{
					\frac{d}{2} + k + m}(px)}{p^{\frac{d}{2} + k - m}}.
		\end{equation*}
		We use the asymptotics of the modified Bessel function 
		\begin{equation*}
			K_{\frac{d}{2} + k + m}(px) = \frac{1}{(px)^{\frac{d}{2} + k + m}}2^{\frac{d}{2} + k + m -1}\Gamma\left(\frac{d}{2} + k + m \right), \qquad x \ll 1.
		\end{equation*}
		This yields
		\begin{multline*}
			\Delta_0(x =0, \gamma, d) = \frac{2}{\pi^{\frac{d}{2}}}\sum_{p=1}^{\infty}\frac{1}{p^d}\sum_{k=0}^{\infty}\frac{1}{k!}\left( -16 \frac{\gamma}{p^2}\right)^k \left( \frac{1}{4}\right)_k \left( \frac{3}{4}\right)_k \sum_{m=0}^{2k}\frac{1}{m!}\frac{\left(-2k \right)_m }{\left(\frac{1}{2} \right)_m}\Gamma\left(\frac{d}{2} + k + m \right)  = \\\frac{2}{\pi^{\frac{d}{2}}}\sum_{k=0}^{\infty}\frac{1}{k!}\left( -16 \gamma\right)^k \zeta(d + 2k) \left( \frac{1}{4}\right)_k \left( \frac{3}{4}\right)_k \sum_{m=0}^{2k}\frac{1}{m!}\frac{\left(-2k \right)_m }{\left(\frac{1}{2} \right)_m}\Gamma\left(\frac{d}{2} + k + m \right).
		\end{multline*}
		We now use the fact that 
		\begin{equation*}
			\Gamma\left(\frac{d}{2} + k + m \right) = \Gamma\left(\frac{d}{2} + k\right) \left(  \frac{d}{2} + k \right)_m = \Gamma\left(\frac{d}{2}\right) \left( \frac{d}{2} \right)_k \left(  \frac{d}{2} + k \right)_m 
		\end{equation*} 
		to  write
		\begin{equation*}
			\Delta_0(x =0, \gamma, d) =    \frac{2}{\pi^{\frac{d}{2}}}\Gamma\left(\frac{d}{2}\right)\sum_{k=0}^{\infty}\frac{1}{k!}\left( -16 \gamma\right)^k \zeta(d + 2k) \left( \frac{1}{4}\right)_k \left( \frac{3}{4}\right)_k \left( \frac{d}{2} \right)_k \sum_{m=0}^{2k}\frac{1}{m!}\frac{\left(-2k \right)_m \left(  \frac{d}{2} + k \right)_m   }{\left(\frac{1}{2} \right)_m}.
		\end{equation*}
	 	 We now notice that
	 	 \begin{equation*}
	 	 	\sum_{m=0}^{2k}\frac{1}{m!}\frac{\left(-2k \right)_m \left(  \frac{d}{2} + k \right)_m   }{\left(\frac{1}{2} \right)_m} = {}_2F_1\left(-2k, \frac{d}{2} + k; \frac{1}{2}; 1\right), 
	 	 \end{equation*}
 	 	which allows us to arrive at the final result (Eq.~(\ref{Delta01}))
 	 	\begin{equation*}
 	 		\Delta_0(x=0, \gamma, d) = \frac{2}{\pi^{\frac{d}{2}}}\Gamma\left(\frac{d}{2}\right)\sum_{k=0}^{\infty}\frac{(-16 \gamma)^k}{k!}\left( \frac{1}{4}\right)_k \left( \frac{3}{4}\right)_k\left( \frac{d}{2}\right)_k \zeta(d + 2k) {}_2F_1\left(-2k, \frac{d}{2} + k; \frac{1}{2}; 1\right).
 	 	\end{equation*} 	 	
		   \subsection{The critical point - truncation}
		We recall the expression for the amplitude at the critical point Eq.~(\ref{Delta01}) 
		\begin{equation*}
			\Delta_0(x=0, \gamma, d) = \frac{2}{\pi^{\frac{d}{2}}}\Gamma\left(\frac{d}{2}\right)\sum_{k=0}^{\infty}\frac{(-16 \gamma)^k}{k!}\left( \frac{1}{4}\right)_k \left( \frac{3}{4}\right)_k\left( \frac{d}{2}\right)_k \zeta(d + 2k) {}_2F_1\left(-2k, \frac{d}{2} + k; \frac{1}{2}; 1\right).
		\end{equation*}
		We want to prove that the $k$-summation truncates in the case of odd dimensionalities. We write that for $d = 2l + 1$ with $l \in \mathbb{Z}_+$
	 	\begin{multline*}
	 		\Delta_0(x=0, \gamma, 2l + 1) = \\  \frac{2}{\pi^{l}}\left( \frac{1}{2}\right)_l \sum_{k=0}^{\infty}\frac{(-16 \gamma)^k}{k!}\left( \frac{1}{4}\right)_k \left( \frac{3}{4}\right)_k\left( l + \frac{1}{2}\right)_k \zeta(2l + 2k + 1) {}_2F_1\left(-2k, l + k + \frac{1}{2}; \frac{1}{2}; 1\right).
	 	\end{multline*}
 		As found in literature \cite{derez}, the hypergeometric function ${}_2F_1$ satisfies
        \begin{equation*}
	   		{}_2F_1\left(a,b;c;z\right) = (1 - z)^{c- (a + b)}{}_2F_1\left(c-a,c-b;c;z\right).
	   	\end{equation*} 
	   	This implies that
            \begin{multline*}
 			{}_2F_1\left(-2k, l + k + \frac{1}{2}; \frac{1}{2}; 1\right) = \lim_{\varepsilon \to 0} {}_2F_1\left(-2k, l + k + \frac{1}{2}; \frac{1}{2}; 1 - \varepsilon\right) = \\  \lim_{\varepsilon \to 0} \varepsilon^{k-l}{}_2F_1\left(2k + \frac{1}{2}, -(l + k); \frac{1}{2}; 1 - \varepsilon\right). 
 		\end{multline*}
 		We have $k, l \in \mathbb{Z}_+$. Hence,  ${}_2F_1\left(2k + \frac{1}{2}, -(l + k); \frac{1}{2}; x\right)$ is a polynomial function of $x$ of degree not higher than $l + k$, which follows from the definition of the hypergeometric series Eq.~(\ref{hyper}). This implies
 		\begin{equation*}
 			\lim_{\varepsilon \to 0} {}_2F_1\left(2k + \frac{1}{2}, -(l + k); \frac{1}{2}; 1 - \varepsilon\right) = c < +\infty,
 		\end{equation*}
 		from which it follows that
 		\begin{equation*}
 			\lim_{\varepsilon \to 0}\varepsilon^{k-l} {}_2F_1\left(2k + \frac{1}{2}, -(l + k); \frac{1}{2}; 1 - \varepsilon\right) = 0, \qquad k>l.
 		\end{equation*}
 		Therefore, the summation in $k$ truncates at $k = l$ and $\Delta_0$ can be cast in the form
 		\begin{multline*}
 			\Delta_0(x=0, \gamma, 2l + 1) = \\  \frac{2}{\pi^{l}}\left( \frac{1}{2}\right)_l \sum_{k=0}^{l}\frac{(-16 \gamma)^k}{k!}\left( \frac{1}{4}\right)_k \left( \frac{3}{4}\right)_k\left( l + \frac{1}{2}\right)_k \zeta(2l + 2k + 1) {}_2F_1\left(-2k, l + k + \frac{1}{2}; \frac{1}{2}; 1\right).
 		\end{multline*}
		
            \subsection{The critical point - conformal transformation}
        To verify the result of the truncated series giving the expression for $\Delta_0$ for $d= 2l + 1$, $l \in \mathbb{Z}_+$, we propose an alternative derivation. First, we rewrite the equation for $\omega_{0,s}\left(T, \mu, D \right)$ Eq.~(\ref{omega1}) and set $\mu = 0$, obtaining
				\begin{equation*}
			\beta \omega_{0,s}(T, D, \mu =0 ) =-\frac{8}{(4\pi)^{\frac{d+1}{2}}}\frac{1}{D^{d-1}}\frac{\lambda_{\tau}^{d-1}}{\lambda_t^{d-1}} I,
		\end{equation*}
		with
		\begin{equation*}
			I:=\sum_{p=1}^{\infty}\frac{1}{p^d}\int_0^{\infty}dt t^{d-1}\int_{\mathbb{R}}dxe^{- \left(x^2 + \frac{\gamma}{p^2}t^2x^4 + ixt \right) }.
		\end{equation*}
		We focus on the case of $d = 2l + 1$. From the series expansion we obtained
		\begin{equation*}
			I =   4^l\pi \left( \frac{1}{2}\right)_l\sum_{k=0}^l\frac{(-16\gamma)^k}{k!} \left( \frac{1}{4}\right)_k\left( \frac{3}{4}\right)_k \left( l + \frac{1}{2} \right)_k\zeta\left(2l+ 2k + 1 \right)  {}_2F_1\left(-2k, l + k + \frac{1}{2}; \frac{1}{2}; 1 \right).
		\end{equation*}
		The series expansion also assumes that we may interchange the summation in $k$ with all the integrations. It is far from obvious whether such an assumption is correct.  We verify the result by performing a conformal transformation of variables
		$$ t \mapsto \tilde {t} = \frac{t}{\sqrt{1 + (tx)^2\frac{\gamma}{p^2}}}, \qquad x \mapsto \tilde {x}  = x\sqrt{1 + (tx)^2\frac{\gamma}{p^2}}.$$
		We note that such a change of variables implies
		  $$ tx = \tilde{t}\tilde{x}, \qquad dtdx= d\tilde{t}d\tilde{x}$$
		and allows us to rewrite the integral in the form
		\begin{equation*}
			I =  \sum_{p=1}^{\infty}\frac{1}{p^{2l + 1}}\int_0^{\infty}dtt^{2l} \int_{\mathbb{R}}dx\left(1 +(tx)^2\frac{\gamma}{p^2} \right)^l e^{- \left(x^2 + ixt \right) }.
		\end{equation*}
		We now use Newton's identity 
		\begin{equation*}
			\left(1 +(tx)^2\frac{\gamma}{p^2} \right)^l = 
			\sum_{k=0}^l\binom{l}{k}\left((tx)^2\frac{\gamma}{p^2} \right)^k
		\end{equation*}
		which leads to
		\begin{equation*}
			I = \sum_{k=0}^l\zeta(2l + 2k +1)\binom{l}{k}\gamma^k\int_0^{\infty}dtt^{2(l + k)} \int_{\mathbb{R}}dx e^{- \left(x^2 + ixt \right) }x^{2k}.
		\end{equation*}
		We use the lemma Eq.~(\ref{ind1})
		\begin{equation*}
			\int_{\mathbb{R}}dx e^{- \left(x^2 + ixt \right) }x^{2k} = \sqrt{\pi}e^{-\frac{t^2}{4}}\left(\frac{1}{2} \right) _k{}_1F_1\left(-k; 
			\frac{1}{2}; \frac{t^2}{4} \right)
		\end{equation*}
		and calculate
		\begin{multline*}
			\int_0^{\infty}dt t^{2(l + k)}e^{-\frac{t^2}{4}}{}_1F_1\left(-k; 
			\frac{1}{2}; \frac{t^2}{4} \right) =\\ 4^{l + k}\int_0^{\infty}dx x^{l + k + \frac{1}{2} -1}e^{-x}{}_1F_1\left(-k; 
			\frac{1}{2}; x \right) = \\ 4^{l + k}\sum_{m=0}^k\frac{1}{m!}\frac{\left(-k \right)_m }{\left( \frac{1}{2}\right)_m }\int_0^{\infty}dx x^{l + k +  m +\frac{1}{2} -1}e^{-x} = \\ 4^{l + k}\sum_{m=0}^k\frac{1}{m!}\frac{\left(-k \right)_m }{\left( \frac{1}{2}\right)_m } \Gamma\left( l + k +  m +\frac{1}{2} \right)  = \\  4^{l + k}\sum_{m=0}^k\Gamma\left( l + k  +\frac{1}{2} \right) \frac{1}{m!}\frac{\left(-k \right)_m }{\left( \frac{1}{2}\right)_m } \left( l + k  +\frac{1}{2} \right)_m  = \\ 4^{l + k}\sqrt{\pi}\left(\frac{1}{2} \right)_l \left(l + \frac{1}{2} \right)_k{}_2F_1\left(-k, l+k+\frac{1}{2};\frac{1}{2};1 \right).
		\end{multline*}
		This leads to the final result
		\begin{equation*}
			I = 4^l \pi \left( \frac{1}{2}\right)_l \sum_{k=0}^{l}(4 \gamma)^k\binom{l}{k} \left( \frac{1}{2}\right)_k \left( l + \frac{1}{2} \right)_k \zeta(2l + 2k + 1) {}_2F_1\left(-k, l + k + \frac{1}{2}; \frac{1}{2}; 1 \right).
		\end{equation*}
		This result is equivalent to the one obtained from the series expansion if $\forall l, k \in \mathbb{Z}_+k \leq l$
		\begin{equation*}
			{}_2F_1\left(-2k, l + k + \frac{1}{2}; \frac{1}{2}, 1 \right) = \frac{\left(-l \right)_k\left( \frac{1}{2}\right)_k  }{4^k\left( \frac{1}{4}\right) _k\left( \frac{3}{4}\right) _k}{}_2F_1\left(-k, l + k + \frac{1}{2}; \frac{1}{2}; 1 \right).
		\end{equation*}
		This identity has been numerically verified for $l \in \{0,1,\ldots, 10\}$ and we leave it as a conjecture.
\section{Correlation Length}
To find the asymptotic $\frac{D}{\sqrt{\beta \tau}}  \gg 1 $ behavior of the surface contribution to the grand canonical potential, we employ the saddle-point method to Eq.~(\ref{saddle2}) and Eq.~(\ref{saddle3}), recalling the latter 
\begin{equation}
	\phi_p\left(t,x\right) = b^3t + b^3 x^2 t + \frac{1}{4}a^4b^3 x^4 t + ipbx,
\end{equation}
with $b = \sqrt{\beta |\mu|}$ and $a = \left(\frac{4\tau' |\mu|}{\tau^2}\right)^{\frac{1}{4}} = \left(\frac{\xi_{4}}{\xi_{2}}\right)$.  \\  
We look for such values of $\tilde{t}, \tilde{x}$ that
\begin{equation}
	\frac{\partial \phi_p }{\partial t}\vert_{\left(t, x\right) = \left(\tilde{t}, \tilde{x}\right)} = \frac{\partial \phi_p }{\partial x}\vert_{\left(t, x\right) = \left(\tilde{t}, \tilde{x}\right)} = 0. 
 \label{saddle4}
\end{equation}   
In the saddle-point approach that the asymptotic $\frac{D}{\sqrt{\beta 
\tau}} \gg 1$ behavior of $I_p$ (Eq.~(\ref{saddle2})) is given by $e^{-\frac{D}{\sqrt{\beta \tau}}\phi_p\left(\tilde{t},\tilde{x}\right)}$, i.e. by replacing the integral with the extreme value of the integrand.\\ 
The condition of vanishing partial derivatives in Eq.~(\ref{saddle4}) gives the system of equations 
\begin{equation}
	\begin{cases}
		1 + x^2 + \frac{1}{4}a^4 x^4 = 0, \qquad \quad \quad (1) \\
		2b^2 xt + a^4b^2 x^3 t = -ip. \qquad (2)
	\end{cases}	
 \label{saddle5}
\end{equation}
Eq.~(\ref{saddle5}.1) has qualitatively different behavior for different values of the parameter $a$.
For $a < 1$ ($\frac{\xi_{4}}{\xi_2} < 1$), its solutions $\tilde{x}$ are such that $\tilde{x}^2$ is real and negative. For $a > 1$ ($\frac{\xi_{4}}{\xi_2} > 1$), its solutions $\tilde{x}$ are such that $\tilde{x}^2$ is purely imaginary. For $a = 1$ ($\frac{\xi_4}{\xi_2} = 1$), the system of equations has no solutions. This indicates that we are unable do derive the asymptotic behavior using the saddle-point method for $a=1$. However,  we will be able to evaluate the asymptotic behavior for $a=1$ by taking $a \to 1^+$ and $a \to 1^-$ limits. \\  In what follows we consider the cases $a < 1$, $a > 1$ and $a = 1$ separately. \\ \\
\underline{\textbf{I. $a < 1$}}
\\  
In the case of $a < 1$ ($\frac{\xi_4}{\xi_2} < 1$), the system of equations has four solutions. However, we disregard two of them, due to the fact that in these solutions $\Re{\tilde{t}} <0$, which means that the corresponding saddle-points lie outside of the relevant integration contour. Indeed, the integration in $t$ is over the set $t > 0$. The other two solutions we analyze in detail. \\ 
The first solution (I.a) with $\Re{\tilde t} > 0$ is given by 
\begin{equation*}
	\begin{cases}
		\tilde{t}^{(a)} = \frac{p}{b^2}\sqrt{\frac{a^4}{8(1-a^4)(1 - \sqrt{1 - a^4})}}, \\
		\tilde{x}^{(a)} = \frac{1}{i} \frac{\sqrt{2}}{a^2}\sqrt{1 -\sqrt{1 - a^4}}, 
	\end{cases}
\end{equation*} 
which gives
\begin{equation*}
	\phi_p^{(\text{I.a})} :=\phi_p\left(\tilde{t}^{(a)},\tilde{x}^{(a)}\right) = bp\frac{\sqrt{2}}{a^2}\sqrt{1 -\sqrt{1 - a^4}} = p\sqrt{\beta \tau}\sqrt{\frac{\tau}{2 \tau'}}\sqrt{1 - \sqrt{1 - \frac{4 \tau' |\mu|}{\tau^2}}}.
\end{equation*}
The second solution (I.b) with $\Re{t} > 0$ is given by
\begin{equation*}
	\begin{cases}
		\tilde{t}^{(b)} = \frac{p}{b^2}\sqrt{\frac{a^4}{8(1-a^4)(1 + \sqrt{1 - a^4})}}, \\
		\tilde{x}^{(b)} = i \frac{\sqrt{2}}{a^2}\sqrt{1 +\sqrt{1 - a^4}}, 
	\end{cases}
\end{equation*} 
which gives
\begin{equation*}
	\phi_p^{(\text{I.b})} :=\phi_p\left(\tilde{t}^{(b)},\tilde{x}^{(b)}\right) = - bp\frac{\sqrt{2}}{a^2}\sqrt{1 +\sqrt{1 - 4a}} = -p\sqrt{\beta \tau}\sqrt{\frac{\tau}{2 \tau'}}\sqrt{1 + \sqrt{1 - \frac{4 \tau' |\mu|}{\tau^2}}}.
\end{equation*}
We argue that we must disregard the second solution (I.b). We note that the resulting $\phi_p$ is unphysical, and the reason for its unphysicality is twofold. If we were to accept this solution, we would obtain that the surface contribution to the grand canonical potential diverges exponentially with increasing distance $D$ between the walls confining the gas. This implies that in this system, the thermodynamic limit does not exists. Also, we expect to observe a power-law divergence of the correlation length as the bulk critical point is approached (in the considered case $\mu \to 0^-$), which implies that in this limit $\phi_p \to 0$. However, one sees that as $\mu \to 0^-$, $\phi_p \to -\sqrt{\frac{\beta \tau^2}{\tau'}}$ holds. On grounds of this physical argumentation, we disregard the solution (I.b).
 \\ \\  We take a closer look at the solution (I.a) and note that asymptotically  for $a \ll 1$, one has 
\begin{equation*}
		\phi_p\left(\tilde{t}^{(a)},\tilde{x}^{(a)}\right) = p \sqrt{\beta \tau}\sqrt{\frac{|\mu|}{\tau}} \left[1 + \frac{1}{2} \frac{\tau' |\mu|}{\tau^2} + \frac{7}{8} \left(\frac{\tau' |\mu|}{\tau^2}\right)^2 + O\left(\left(\frac{\tau' |\mu|}{\tau^2}\right)^3\right)\right].
\end{equation*}
We identify $\phi_p = \sqrt{\beta \tau} \frac{p}{\xi}$ and state that the leading order in $|\mu|$ of the inverse of correlation length $\frac{1}{\xi}$ is equal to $\frac{1}{\xi_2}$. The higher order terms are expressed in terms of a series expansion in $\frac{\tau' |\mu|}{\tau^2} = \frac{a^4}{4}$. This shows that in the case of the mixed dispersion, we obtain the same critical exponent $\nu = \frac{1}{2}$, as in the case of purely quadratic dispersion. Moreover, as $\tau'$ is tuned to $0$, the correlation length of the system with quadratic dispersion is recovered in a continuous manner. Also, if one were to calculate the saddle-point solution for the gas with the quadratic dispersion, one would find that $\tilde{t}^{(a)} \to \tilde{t}_{2}$ and $\tilde{x}^{(a)} \to \tilde{x}_{2}$ as $\tau' \to 0$, where $\tilde{t}_{2}$ and $\tilde{x}_{2}$ are the solutions of the saddle-point equation with $\tau' = 0$.  \\ \\ 
\underline{\textbf{II. $a >1 $}} \\ 
In the case of $a > 1$ ($\frac{\xi_4}{\xi_2} > 1$), the system of equations once again has four solutions. As before, we disregard two of them, in which $\Re{\tilde{t}}<0$. \\  This leaves two solutions, the first (II.a) given by
\begin{equation*}
	\begin{cases}
		\tilde{t}^{(a)} =  \frac{ip}{2\sqrt{2}b^2} \frac{a}{\sqrt{a^4 -1}}e^{-\frac{i}{2}\arctan{\sqrt{a^4 - 1}}}, \\ 
		\tilde{x}^{(a)} = \frac{1}{i}\frac{\sqrt{2}}{a}e^{\frac{i}{2}\arctan{\sqrt{a^4 - 1}}},
	\end{cases}
\end{equation*}
where $\arctan(x)$ is the inverse tangent function. This solution leads to 
\begin{equation*}
	\phi_p^{(\text{II.a})} :=\phi_p\left(\tilde{t}^{(a)},\tilde{x}^{(a)}\right) = \frac{\sqrt{2}bp}{a} e^{\frac{i}{2}\arctan{\sqrt{a^4 - 1}}} = p\sqrt{\beta \tau}\left(\frac{|\mu|}{\tau'}\right)^{\frac{1}{4}}e^{\frac{i}{2}\arctan{\sqrt{\frac{4 \tau' |\mu|}{\tau^2} - 1}}}.
\end{equation*}
The second solution (II.b) is given by
\begin{equation*}
	\begin{cases}
		\tilde{t}^{(b)} = \frac{1}{i}\frac{p}{2\sqrt{2}b^2} \frac{a}{\sqrt{a^4 -1}}e^{\frac{i}{2}\arctan{\sqrt{a^4 - 1}}}, \\ 
		\tilde{x}^{(b)} = \frac{1}{i}\frac{\sqrt{2}}{a}e^{-\frac{i}{2}\arctan{\sqrt{a^4 -1}}},
	\end{cases}
\end{equation*}
which gives 
\begin{equation*}
	\phi_p^{(\text{II.b})} :=\phi_p\left(\tilde{t}^{(b)},\tilde{x}^{(b)}\right) = \frac{bp}{a^{\frac{1}{4}}} e^{-\frac{i}{2}\arctan{\sqrt{4a - 1}}} = p\sqrt{\beta \tau}\left(\frac{|\mu|}{\tau'}\right)^{\frac{1}{4}}e^{-\frac{i}{2}\arctan{\sqrt{\frac{4 \tau' |\mu|}{\tau^2} - 1}}}.
\end{equation*}
We note that the solutions (II.a) and (II.b) are such that  $\phi_p^{(\text{II.a})}$ is the complex conjugate of ${\phi_p^{(\text{II.b})}}$. This implies that they have the same absolute value and both have to be taken into account in the saddle-point approach. The asymptotic behavior of the integral $I_p$ will be therefore given by the sum of the extreme values of the integrand, which concludes the derivation for Eq.~(\ref{ocs1}) and Eq.~ \ref{corr2}).
\\ \\ 
\underline{\textbf{III. $a = 1$}} \\
The system of equations Eq.~(\ref{saddle5}) has no solutions for $a = 1$. However, we notice that for the solution (I.a), which is valid for $a<1$, and solutions (II.a) and (II.b), which are valid for $a > 1$, we have
\begin{equation}
	\lim_{a \to 1^-}\phi_p^{(\text{I.a})} = \lim_{a \to 1^+}\phi_p^{(\text{II.a})} = \lim_{a \to 1^+}\phi_p^{(\text{II.b})} = bp\sqrt{2}.
\end{equation}
Moreover, we may interpret that the solution $\tilde{t}^{(\text{I.a})}$ as $a \to 1$ diverges to infinity and then, as $a$ increases, its absolute value decreases and it splits into solution $\tilde{t}^{(\text{II.a})}$ and $\tilde{t}^{(\text{II.a})}$. Similarly,  $\tilde{x}^{(\text{I.a})}$ splits into $\tilde{x}^{(\text{II.a})}$ and $\tilde{x}^{(\text{II.a})}$, this time, however, without diverging to infinity at $a=1$. \\ \\
To summarize, we arrived at the result that for $x = \frac{D}{\xi_2}$ and $a=\frac{\xi_4}{\xi_2}$, the asymptotic $x \gg 1 $ behavior of $I_p$ is given by
\begin{equation*}
	I_p \sim \begin{cases}
		e^{-\sqrt{2}p\frac{x}{a^2}\sqrt{1 - \sqrt{1 - a^4}}}, \qquad \qquad \qquad \qquad \qquad \qquad \qquad \qquad \qquad \quad   a<1,  \\  e^{-\sqrt{2}px}, \qquad \qquad \qquad \qquad \qquad \qquad \qquad \qquad \qquad \qquad \qquad  \qquad a=1,  \\ e^{-\sqrt{2}p\frac{x}{a}\cos{\left(\frac{1}{2} \arctan{\sqrt{a^4 - 1}}\right)}}\cos\left(\sqrt{2}p\frac{x}{a}\sin{\left(\frac{1}{2} \arctan{\sqrt{a^4 - 1}}\right)} \right), \quad a > 1. 
	\end{cases}
\end{equation*}
 We arrived at the expression for the correlation length 
\begin{equation*}
	\xi^{-1} = \begin{cases}
		\sqrt{\frac{\tau}{2\tau'}}\sqrt{1 - \sqrt{1 - \frac{4\tau' |\mu|}{\tau^2}}}, \quad \qquad \qquad \quad\frac{4 \tau' |\mu|}{\tau^2} <1,\\  \sqrt{\frac{2 |\mu|}{\tau}} = \sqrt{\frac{\tau}{2 \tau'}},\qquad \qquad \qquad \qquad \qquad \frac{4 \tau' |\mu|}{\tau^2} = 1, \\  \\
		\left(\frac{|\mu|}{\tau'}\right)^{\frac{1}{4}}\sin{\left(\frac{1}{2}\arctan{\sqrt{\frac{4\tau' |\mu|}{\tau^2} -1}}\right)}, \qquad \frac{4 \tau' |\mu|}{\tau^2} = 1. 
	\end{cases}
\end{equation*}
\end{document}